\documentclass{sigcomm-alternate}
\usepackage{graphicx}
\usepackage{url}
\usepackage{booktabs}
\usepackage{subfigure}
\usepackage{color}

\begin{document}

\title{Taming Limits with Approximate Networking}
\author{\large Junaid Qadir$^{1}$, Arjuna Sathiaseelan$^{2}$, Liang Wang$^{2}$, Jon Crowcroft$^{2}$\\
\normalsize $^{1}$Information Technology University (ITU)-Punjab, Lahore, Pakistan\\
\normalsize $^{2}$Computer Laboratory, University of Cambridge, United Kingdom\\
\normalsize Email: junaid.qadir@itu.edu.pk; arjuna.sathiaseelan@cl.cam.ac.uk; liang.wang@cl.cam.ac.uk; jon.crowcroft@cl.cam.ac.uk}
\maketitle

\begin{abstract}


Internet is  the linchpin of modern society, which the various threads of modern life weave around. But being a part of the bigger energy-guzzling industrial economy, it is vulnerable to disruption. It is widely believed that our society is exhausting its vital resources to meet our energy requirements, and the cheap fossil fuel fiesta will soon abate as we cross the tipping point of global oil production. We will then enter the long arc of scarcity, constraints, and limits---a post-peak ``long emergency'' that may subsist for a long time. To avoid the collapse of the networking ecosystem in this long emergency, it is imperative that we start thinking about how networking should adapt to these adverse ``undeveloping'' societal conditions. We propose using the idea of ``\textit{approximate networking}''---which will provide \textit{good-enough} networking services by employing \textit{contextually-appropriate} tradeoffs---to survive, or even thrive, in the conditions of scarcity and limits. 

\end{abstract}

\section{Introduction}

Human beings today enjoy a standard of living unparalleled in human history. This age of abundance is driven largely by technology and in particular through information and communication technology (ICT). The Internet---which has impacted all facets of human life (business, governance, education, leisure) through its ability to connect people and facilitate communication---is widely believed to be a gateway to human prosperity and opportunities. 

But due to the non-sustainable\footnote{The sustainability of a system such as the Internet measures its capacity to endure diverse exogenous pressures (such as resource depletions) while remaining productive indefinitely.} trajectory adopted by the industrialized civilization, the ICT-enabled societal progress has been achieved at a big cost. The fundamental physical limits of a finite world are being taxed by, and will inevitably stall, the exponential trends displayed in human/ society demographics (e.g., population) and ICT (e.g., Moore's law, big data). It is now widely believed that modern civilization is close to exhausting non-trivial resource limits (such as the non-renewable fossil fuels \cite{heinberg2005party}). As this depletion will start to show its effects in the post-peak environment, even developed countries will face degradation of their social and economic systems due to societal collapse (and will ``undevelop'' \cite{Net4Undeveloping}---i.e., despite having some infrastructure, these countries will regress and will become economically and politically unstable).


\subsection{Post-Peak Future of Limits and Scarcity}

A characeristic reason for the ``\textit{collapse}''\footnote{Collapse happens over a long time---decades or sometimes even centuries and should not be confused with an apocalypse that occur suddenly and instantaneously.} of a society is that its citizens commit ``\textit{ecocide}''---i.e., they overuse and exhaust their vital resources \cite{diamond2005collapse}. The industralized world may have also committed an ecocide through its overreliance on fossil fuels. Fossil fuels comprising oil, coal, and natural gas together account for approximately 80\% of global energy consumption, with oil servicing the major bulk \cite{raghavan2011networking}. Many experts are predicting that the production of oil---which is the foundation of the industrial system---will deplete in the near term \cite{hirsch2010peaking}. After the peaking of the world's oil production, the decline of oil production will likely lead to a ``long emergency'' whose effects will be wide reaching and long lasting (that may last for decades). 

A US government commissioned study (called the Hirsch report) attempted to analyze the timing and consequences of oil peaking---the results indicated that peak oil production will occur by 2016. Similar studies elsewhere indicate that this peak may already be behind us---although oil reserves and production capacity are often closely guarded secrets, and the collapse consequences of the post-peak may take some time to manifest itself, some of the signs (e.g., global climatic change, economic slowdown) have already started to emerge. Since fossil-fuels are non-renewable (i.e., they exist in finite non-replenishable quantity), a point will inevitably come where the rate of resource extraction peaks.\footnote{Some experts indicate that we may already be in this post-peak era \cite{raghavan2011networking}.} While there is some ongoing work on using renewable energy for networking \cite{mineraud16}, it is doubtful,  extrapolating current trends, that we will be able to match our inflated requirements with renewable energy in the short term. Such a likely persistent decline in energy will provide a permanent shock to the energy-guzzling industrial ecosystem which will likely lead to a societal collapse. 

It is instructive to note some other reasons for societal collapse noted in literature \cite{diamond2005collapse}, such as: (1) reliance on trading partners; (2) self-inflicted environmental damage; and (3) inflexibility of institutions when change is needed. The latter two reasons are directly relevant to our subject topic. The first reason is also related but indirectly: it shows how dependence on entities can make systems less resilient to disruptions. To be sure, the emergence of an  \textit{an oil-depleted and post-peak future poses fundamental constraints and limits on the Internet architecture and infrastructure}.



\subsection{Networking in the Long Emergency}


To cope with the impending sudden and potentially long-lasting energy shock, it is necessary for the networking architecture to adapt. The key questions in the networking context are: 

\begin{enumerate}

\item \textit{how should we adapt Internet technology so that it becomes sustainable?}; 

\item \textit{how will the Internet users and applications fare if the Internet turns out not to be sustainable?}. 

\end{enumerate}

The answers to these questions can motivate the adaptation of our network architecture and applications so that they become ``\textit{collapse compliant}'' \cite{tomlinson2012collapse}. 

Researchers have only started to look at how should networking adapt to deal with a ``long emergency'' that can arise in an undeveloping environment. In the seminal paper that looked at networking issues for the long emergency \cite{raghavan2011networking}, a number of premises or assumptions were stated. It was assumed that energy and financial constraints will be non-uniform in different regions and will iterate between contraction/ partial recovery but with an overall downward trend. It was also assumed that economic decline will also be a major challenge for networking (apart from the direct impacts of energy scarcity). It was also assumed that user bases will likely shrink as there may be a decrease in the overall use of computing to meet societal needs (and non-digital alternatives may be increasingly deployed). 



\subsection{The Approximation Tradeoff}

\vspace{1mm}
``\textit{Although this may seem a paradox, all exact science is dominated by the idea of approximation}.''---Bertrand Russell
\vspace{1mm}

We necessarily employ approximation in many technologies and sciences. We use approximation in measurement and in digital computing. We use approximation when the problem is too intractable to solve optimally: in such cases, we lower down our targets to \textit{satisficing} (i.e., producing ``good enough'' answers) rather than optimizing \cite{stirling2003satisficing}. 

While ideally speaking, we will like an Internet that is perfect, and has extremely high capacity, bandwidth, and reliability in addition to extremely low or negligible delays, errors, and congestion.\footnote{For practical purposes, the modern fiber-based broadband high-speed networks available in select places (mostly in advanced countries) come close to this ideal.} We call such networks ``ideal networks''. In contrast, we consider ``approximate networks'' that are networks that make some design tradeoffs to deal with varying levels of challenges and impairments. We note here that ideal networks and approximate networks do not define a binary divide but a spectrum of options. We can also define approximate networks as networks that come close to ideal networks in quality, nature, and quantity\footnote{Oxford Dictionary: Approximate (v): come close or be similar to something in quality, nature, or quantity.}. We need ``approximate networking'' when the imperfections of the real world preclude an ``ideal networking'' solution. In particular, an approximate networking solution is appropriate when any of the ideal networking  assumptions---e.g., that there is 24x7 connectivity; an end-to-end path is always available; the end-to-end delay, and the link propagation delay, is never too high (i.e., is less than (half) a second); the networks should not be congested; the networks should not have high error rates---are not met. 

Approximate networking is inspired in part from the emerging architectural trend of ``approximate computing'' \cite{han2013approximate} in which approximations are performed at the hardware level to boost the energy efficiency of systems. Broadly speaking, approximate computing leverages the capability of many computing systems and applications to tolerate some loss of quality and optimality by trading off ``precision'' for ``efficiency'' (however, these may be defined). Approximate networking, in a similar vein, enables a network architecture that allows networking protocols and applications to trade off service quality for efficiency in terms of cost/ affordability/ accessibility. Approximate networking is closely aligned to the philosophy of ``appropriate technology'' since it aims to match the user and the need in complexity and scale. Appropriate technologies are defined as ``small scale, energy efficient, environmentally sound, labor-intensive, and controlled by the local community'' \cite{hazeltine1998appropriate}. Approximate networking also fits well as a collapse informatics solution since it can be used as a ``tool for the study, design, and development of sociotechnical systems in the abundant present for use in a future of scarcity'' \cite{tomlinson2012collapse}. 

\subsection{Comparison of ICTD and LIMITS}

While there is some overlap in the focus of ICT for development (ICTD) and
the research on the problems of the undeveloping world with limits (we refer to this setting in this paper as LIMITS), these areas are fundamentally different. A number of standard ICTD assumptions \cite{brewer2005case} (e.g., Moore's law will hold into the future; there will be increased access to capital and improved business environments; the use of widespread ICT is desirable\footnote{In fact in the context of LIMITS, less ICT may be more desirable than more---e.g., previous research has focused on developing self-obviating systems designed to make themselves superfluous through their use so that ICT reliance is reduced \cite{tomlinson2015self}.}) may be invalid in the context of LIMITS. In addition, in contrast to the ICTD research that emphasizes a developing/developed world split, \textit{the limits considered in our context are more global}. Thus it is anticipated that virtually all nations will experience a reduction in their material standards of living with developed countries also facing the crunch (perhaps even more so since they are more heavily reliant on ICT and infrastructure). The ``\textit{undeveloping countries}'' focused on in the LIMITS context is a generalization of the narrow class of developing regions (focused mostly in ICTD research \cite{brewer2005case}) to also include developed countries that have some infrastructure but also a regressing economical and political climate.

For sure, there can be significant diffusion and reuse of ideas across LIMITS and ICTD. LIMITS solutions can produce innovations that are broadly useful  for localized collapse solutions, emergency response, and ICTD \cite{tomlinson2012collapse}. Similarly, the ICTD research focusing on dealing with scarcity will be valuable in the LIMITS context. The technnique of approximate networking is relevant for both ICTD and LIMITS research. Approximate networking can be used in the LIMITS settings to wean our reliance on energy-inefficient overengineered ``perfect'' ICT while still attaining contextually appropriate service. 

\subsection{Contributions of this paper}

The main contribution of this paper is to position approximate networking as a suitable framework for systematically thinking about networking tradeoffs that will inevitably arise in the era of post-peak world burdened with limits and societal collapse. The resource crunch in such an environment will necessitate a move away from overengineered ``perfect products'' towards \textit{contextually-appropriate} ``good enough'' solutions. The challenge entailed in approximate networking is in determining what these context-appropriate network tradeoffs should be. Approximate networking is relevant both for ICTD research (that focuses on developing countries) and for research in the LIMITS context \cite{Net4Undeveloping} \cite{raghavan2011networking} that focuses more broadly on ``undeveloping countries'' that have regressed due to global limits. 

\section{Why Use Approximate Networking under LIMITS?}

\vspace{1mm}
``\textit{And so we turn to the essentials of our future. In order: food, energy, and--yes---the Internet}''---McKibben.
\vspace{1mm}

In the post-peak era of decline of the industrial society, a number of needs such as food and transportation will become prioritized, but Internet will also likely be a key resource for the post-peak future and thus developing solutions to retain its functionality (even if certain approximation tradeoffs are adopted) will be essential. In this era of scarcity, providing ``ideal networking'' service will not be economically feasible and some sort of tradeoff would be inevitable and an inescapable recourse for network designers. This is not new since we have learnt through decades of experience with the Internet that invariably there are Internet design tradeoffs and there is no single one-size-fits-all solution.  Approximate networking is a useful way to think of tradeoffs that users and applications should consider for sustainable and collapse-compliant networking in the grim situation where we will run out of many of the essential resources (such as cheap energy through fossil fuels). 

Some important reasons we should seriously consider approximate networking for dealing with limits are described next.

\subsection{Coping With Resource Constraints}

In many developing parts of the world, resource constraints (such as limited power, unstable government) are a norm of life. Even at a global level, it is anticipated that the modern fossil fuel based industrial system is not sustainable, and the impending depletion of these resources will probably give rise to a sudden and permanent shock that may lead to economic instability and infrastructural challenges \cite{Net4Undeveloping}. Such a severe permanent energy crisis can have far-reaching consequences on the economy and lead developed countries towards being  ``undeveloping countries'' \cite{Net4Undeveloping}. Approximate networking insights can be used to reorient the design of the Internet's algorithms, protocols, and infrastructure to better manage the overarching energy, societal, material, and economic limits that this looming scarcity-based future will impose.




\subsection{Need of Energy Efficiency}

Information and communication technology (ICT) is a big consumer of world's electrical energy, using up to 5\% of the overall energy (2012 statistics) \cite{Gelenbe2015}. The urgency of developing an energy efficiency manifesto is reinforced when we consider the impending decline of non-renewable energy resources as well as the increased demand of ICT (as more and more people get online and use ICT for exchanging greater and greater amounts of data traffic) \cite{raghavan2011networking}. This strongly motivates the need for energy efficient Internetworking  \cite{raghavan2012intermittent}. The approximate networking trend can augment the hardware-focused approximate computing trend to ensure that the energy crisis is managed through the ingenuous use of approximation. 

\subsection{The Pareto Principle (80-20 Law): \\The Power of ``Good Enough''}

\vspace{1mm}
``\textit{Among the factors to be considered there will usually be the vital few and the trivial many.}''---Turan.
\vspace{1mm}

To help manage the approximate networking tradeoffs, it is instructive to remember the \textit{Pareto principle}, alternatively called the \textit{80-20} rule \cite{koch201180}---which states that roughly speaking that 20\% of the factors result in 80\% of the overall effect. This principle has big implications for approximate networking since this allows us to provide adequate fidelity to ideal networking by only focusing on the most important 20\% of the effects. Alternatively put, this theory states that 80\% of what goes into creating the ideal networking experience provides little cosmetic benefits to the user. The key challenge in approximate networking then becomes the task of determining these all-important essential non-trivial factors. Through this exercise, we can create ``ideal networking'' solutions by identifying which factors are costly to implement but provide little gains allowing these resources to be used more efficiently.  
 
\section{Implementing Approximate Networking under LIMITS}

In this section, we propose some concrete building blocks, or loosely speaking principles, that can support approximate networking solutions. We think that the following 5 principles can be useful for implementing approximate networking under limits: (1) \textit{adopt context-appropriate tradeoffs}; (2) \textit{adopt resource pooling \& bottom-up networking}; architect a (3) \textit{failure-cognizant network design} and (4) \textit{scarcity-inspired network design}; and finally (5) \textit{design for intermittency}. These principles are discussed in turn next.

In deriving these basic building blocks, we have utilized insights from previous works that have proposed principles for computing within limits \cite{raghavan2012intermittent} \cite{raghavan2015abstraction} \cite{chen2015computing}, robust networking \cite{anderson2003design},  and frugal innovation under scarcity and austerity  \cite{radjou2012jugaad} \cite{mullainathan2013scarcity}). 

\subsection{Context-Appropriate Tradeoffs}

\vspace{1mm}
``\textit{Wisdom is intelligence in context}.''---Unknown.
\vspace{1mm}

A \textit{tradeoff} refers to the fact that a design choice can lead to conflicting results in different quality metrics. The performance of computers networks depends routinely on multiple parameters. Since these multiple objectives often conflict with each other, it is rare to find one-size-fits-all solution and tradeoffs have to be necessarily employed. We can borrow concepts from economics to study scarcity and choice. The concept of \textit{opportunity cost}---which is the ``cost'' incurred by going with the current choice and not adopting any other choice---is a key idea that can be used to ensure efficient usage of scarce resources. Another important concept is that of \textit{Pareto optimality}, which refers to a state of resource allocation in which it is not possible to make any one individual better off without making at least one individual worse off. We can make a Pareto improvement, if we can make at least one individual better off without making any other individual worse off. 

\subsubsection{Performance vs. Cost Efficiency}

We can tradeoff performance (measured in metrics such as \textit{resilience, reliability, throughput}) to gain on cost efficiency. 
It is said that one of the easiest way to gain cost efficiency is often by sacrificing resilience and reliability (by employing lesser redundancy) \cite{raghavan2012intermittent}. The catch is that by provisioning a lesser-resourced inexpensive network, we are compromising with capacity and will thus have lower throughput for user applications. It is also worth noting that current networks are optimized to comply to very high standards in terms of high-aiming service level agreements (SLA) that aim for very high availability (e.g., 99.999\%). By considering failure as an option \cite{rice2008failure}---i.e., by allowing some failures, and not trying to eliminate them completely at a very high cost---future networks can become significantly more cost efficient. 

\subsubsection{Coverage vs. Consumed Power}

In networking, there is often a direct relationship between coverage and consumed power: typically, higher-powered transmissions have a large coverage range. Approximate networking can improve the energy efficiency of systems by incorporating  intermittency as a degree of freedom for controlling the consumed power. Since nodes do not need to communicate at all times, researchers have proposed putting to sleep parts of the infrastructure---such as the base transceiver station (BTS) of cellular systems \cite{heimerl2013experiment}---to save on energy costs. The research challenge that arises from this approach is to ensure that the infrastructure can be activated when communication is needed and no messaged or lost or delayed inordinately. 


\subsubsection{Performance vs. Coverage/ Reliability}

In certain cases, it may be appropriate to tradeoff coverage for performance, while the opposite may be true in other situations. In wireless networks, there is a tradeoff between the throughput and the coverage (and the reliability) of a transmission---i.e., for higher-rate transmissions, the coverage area is typically smaller, and the chances of bit errors higher. 

\subsubsection{Managing the tradeoffs in networking}

While we have described the main tradeoffs involved in approximate networking and have discussed how they may be visualized, the all-important question still remains to be addressed: \emph{How can we effectively manage these approximate networking tradeoffs?} This is very much an open issue and some open important questions regarding tradeoffs are as follows: 

\begin{enumerate}

\item \textit{How do we quantify when our approximation is working and when it is not?} 

\item \textit{How do we measure success in managing the service quality/ accessibility tradeoff?} 

\item \textit{How do we measure the cost of approximation in terms of performance degradation?}

\item \textit{How to dynamically control the approximation tradeoffs according to the network condition.}

\item \textit{How to also incorporate social optimality into a user's  approximation decision? (A selfish use of approximate networking can improve one user's performance at the cost of all others.) }

\item \textit{How to design proper incentivizes for the service provider and the user so that both act harmoniously for provisioning a  customer-centric contextually-appropriate service?}

\item \textit{How to outsource some of things that we do on the current Internet to less-costly and less energy-intensive offline methods (while ensuring that we get enough QoS that is necessary for our applications) \cite{greer2008}?} 

\end{enumerate}

\subsection{Resource Pooling \& Bottom-Up Networking} 

\vspace{1mm}
``\textit{Innovative bottom-up methods will solve problems that now seem intractable---from energy to poverty to disease}.''---Vinod Khosla
\vspace{1mm}


Broadly speaking, resource pooling involves aggregating a collection of networked resources such that they behave collectively as a single unified resource pool and developing mechanisms for shifting load between the various parts of the unified resource pool. The main benefits of resource pooling include greater reliability and increased robustness against failure; better ability to handle surges in load on individual resources; and, increased utilization \cite{wischik2008resource}. Resource pooling is well suited in scarcity-afflicted approximate networking settings, where maintaining dedicated IT infrastructure and staff is especially cost prohibitive for small-scale entrepreneurs, business owners, and non-profits. Resource pooling can also be especially influential in a world with limits since resource pooling naturally allows some slack in dealing with with scarcity and failures. 


\subsubsection{Encouraging Versatility, Recombination, and Reuse}

The Latin word versatilis connotes turning, or having the capable of turning to varied subjects or tasks. In networks burdened with limits, it will be important to reuse existing resources versatilely for various new settings that may arise. It is possible that a future Internet may require internetworking of partially-connected networks using totally different locally-optimal protocol stacks \cite{raghavan2011networking}. A good example of a versatile approximate networking technique is the use of software defined radio (SDR). SDRs by their versatile nature are radio chameleons that can use software programmability to run completely different protocols at different times  (e.g. CDMA and Wi-Fi). It will also be important for networks operating under limits to maximize their efficiency by avoiding the waste of resources. In this regard, we can deploy \textit{dynmamic spectrum access} \cite{zhao2007survey} to provide secondary users (SUs) access to a primary network when it is not being used by the licensed primary users (PUs). We can also deploy \textit{scavenger transport protocols} to provide less-than-best-effort (LBE) service \cite{ros2013less}. LBE service can be used to facilitate background applications in accessing the unused capacity of backhaul links without impacting the performance of priority applications.

\subsubsection{Community/ Crowdsourced/ DIY networks}

Being a bottom-up cost-effective approach for building networks, community networking is especially promising for approximate networking under limits. Community networks can be used to implement efficient usage of resources through better resource sharing. In recent times, it has even become possible to develop community cellular networks using low-cost software defined radios (SDRs) and open-source software such as OpenBTS \cite{heimerl2013local}. Such community-driven projects can be used to provide approximate networking services where traditional ideal networking solutions are not feasible. Community networks can allow inclusive services in the future of limits by providing non-priority users (e.g., underprivileged users in developing regions) LBE access to networking services while also ensuring appropriate quality of service (QoS) for priority users (such as the users who are contributing their own networking infrastructure) \cite{sathiaseelan2013lcd}.

\subsubsection{Multiplicity, Redundancy, and Slack}

The principles of multiplicity, redundancy, and using slack may look out of place in a paper on approximate networking. But these ideas are somewhat counterintuitively quite important under limits since any approximate networking solution for such environments that does not have redundancy and slack inbuilt will be debilitatingly fragile. It has been shown in literature that keeping a margin or keeping some slack is a key to frugal innovation \cite{radjou2012jugaad} and thriving in scarcity-afflicted environments \cite{mullainathan2013scarcity}. Approximate networks can also exploit the power of multiplicity by supporting heterogeneous technologies and by resource pooling a diverse collection of paths and thereby unlock the inherent redundancy of the Internet. In particular, approximate networking solutions can leverage the inherent diversity and multiplicity of networks to reap the benefits of increased reliability, efficiency, and fault tolerance \cite{qadir2015exploiting}. 






%


\subsection{Failure Cognizant Network Design} 

\vspace{1mm}
``\textit{Hoping for the best, prepared for the worst, and unsurprised by anything in between.}''---Maya Angelou.
\vspace{1mm}

\subsubsection{Design for Failure/Collapse} 


Approximate networking solutions must be designed assuming an inevitable presence of failures/ weaknesses/ deficiencies. The applications must then be designed to withstand and cope with some failures while still proving ``good enough'' services. This is necessary since in the LIMITS setting, we will plausibly deal with many impairments such as long signal propagation delays, high bit error rates (BER), frequent disruptions and unstable or intermittently available links; high congestion; very low data rates; and variable bandwidth. It will be helpful to plan for such a state by assuming severe resource deficiencies such as 25\% less power; 25\% less connectivity; 10x more volatility; 10x more failures; 10x less non-renewable materials; 10x greater variation in societal structure \cite{raghavan2012intermittent}. Failures should be anticipated and even intentionally utilized where appropriate---e.g., it may be helpful to intentionally cause errors or minor failures; the ``random early detection'' (RED) congestion control algorithm intentionally drops some packets when the average queue buffer lengths are more than a threshold (e.g. 50\%) to signal implicitly to the sender the rising congestion. 

\subsubsection{Robust Design For Avoiding Failure/Collapse}

The principle about designing for failure should not be construed to mean that approximate networking should not try to avoid failure. To the contrary, it is very important for approximate networking solutions to be robust\footnote{We define some property of a system to be robust if it is invariant with respect to some set of perturbations. Fragility is defined as the opposite of robustness.}  and to fail gracefully when subsystems fail. Approximate networking solutions should aim to avoid disruption due to failures by adopting robust tradeoffs that make the solution failure proof or resilient. Towards this end, it has been pointed out in literature that the solutions should ``keep the margin'' \cite{radjou2012jugaad} and should have ``spare bandwidth'' \cite{mullainathan2013scarcity} when working in scarcity-afflicted environments. 

Approximate networking solutions will also do well to incorporate insights about robust networking gleaned from robust highly-evolved biological and technological systems (such as the Internet)  that gracefully degrade when afflicted with failures. Alderson \& Doyle \cite{alderson2010contrasting} have argued that complexity arises in such systems in order to provide robustness to uncertainty in their environments---however, this complexity can also be a source of fragility, leading to a ``robust yet fragile'' tradeoff in system design. The need to scale out at all levels of the architecture---at both the level of distributed systems and at the macro system level---is also emphasized in the paper by Bhargavan \cite{raghavan2015abstraction} for creating ``benign systems'', which are computer systems that are less likely to produce harmful impacts to the ecosystem and society. 


\subsubsection{Decentralization}

We've earlier discussed how approximate networking should adopt the principles of appropriate technology such as building systems that are simple, locally reproducible, composed of local materials and resources. More generally, in a world burdened by limits, we would like to have resilient infrastructures, and one way of building resilience is to adopt decentralized architectures. Decentralized architectures are more resilient since they can more easily absorb change and disturbance. In a previous work \cite{tomlinson2015toward}, the authors recommend meeting critical human needs such as food, water, energy, communications using alternative decentralized infrastructures (ADIs), which comprise coordinated distributed collections of small-scale systems and services (in preference to large centralized interdependent critical infrastructures that are used in today's settings of abundance and stability). The trend of edge computing, another instance of decentralized computing in which computing is not performed in a centralized cloud but at the edge close to the user, can also be useful in the LIMITS settings \cite{sathiaseelan2015scandex}. 

\subsubsection{Design for Intermittency}

In the future world taxed by limits, it will be not be possible, nor desirable, to provide networking service all the time. Approximate networking should also be designed while accounting for  intermittent availability of resources. This intermittent access may be enforced or volitional. As an example of intentional use of intermittency (also called \textit{``duty cycling''}), we note that the technique of volitional intermittency can be leveraged in a context-appropriate fashion to provide satisfactory QoS while also saving on energy costs. Approximate networking can draw insights from the rich literature on \textit{delay-tolerant networking} (DTN) \cite{fall2003delay} and \textit{opportunistic networking} \cite{huang2008survey} that are also focused on disrupted and intermittently accessible networks. 

Traditional cellular networks are designed mostly for performance goals and are not optimized for energy. It has been shown that the ``always-on’’ service approach of traditional cellular networks results in an energy consumption profile that is agnostic of load (i.e., the energy consumption is wastefully high even for light loads) \cite{mohamed2015control}. Such cellular networks can benefit greatly by purposefully employing the approximate networking techniques of intermittency and duty cycling. Such networks can also leverage new advancements in cellular infrastructures---such as the control-data separation architecture for cellular radio access networks \cite{mohamed2015control}---for implementing approximate networking more flexibly. 





\subsection{Scarcity Inspired Network Design}

\vspace{1mm}
``\textit{What really makes it an invention is that someone decides not to change the solution to a known problem, but to change the question.}''---Dean Kamen
\vspace{1mm}

Approximate networking can benefit from the insights presented in \cite{radjou2012jugaad}, in which the following guiding principles were
provided for frugal innovation in in complex challenging scarcity-afflicted settings: (1) \textit{Seek opportunity in adversity}; (2) \textit{Do more with less}; (3) \textit{Think and act flexibly}; (4) \textit{Keep it simple}; and (5) \textit{Include the margin}. 

\subsubsection{Protocols and Services Optimized for Scarcity}

In the future world of scarcity, the majority of the people may be encumbered by poor network connectivity, and prohibitively slow/ unstable services, that are ill suited to the design of conventional protocols and services \cite{chencomputing}. This motivates the development of optimized protocols and services that can work well in such poor-connectivity scenarios. We motivate this by discussing transport-layer protocols (while noting that scarcity-aware protocols are needed at all layers). This is well known in graph theory and network science 

When faced with a severely congested links (which are not uncommon in challenged environments), the congestion control service of conventional Internet transport protocols such as TCP starts to break down as the system tends towards \textit{sub-packet regimes}, where a typical per-flow throughput becomes less than 1 packet per round-trip time (RTT). In small packet regimes, the performance of TCP degrades resulting in severe unfairness, high packet loss rates, and stuttering flows due to repetitive timeouts. For such sub-packet regimes, innovative active queue management (AQM) solutions can be deployed to reduce timeouts and thereby improve fairness and performance predictability \cite{chen2014taq}.





\subsubsection{Simple Approximate Networking Solutions}

Simplicity when coupled with convenience and accessibility can result in wide adoption of approximate networking as it has been shown time and again that users are willing to trade off fidelity of user experience to gain on accessibility and convenience. Simplicity has always been considered a virtuous design trait---e.g., this has been codified in the engineering principles of KISS (``Keep it Simple, Stupid'') and the ``Occam's Razor'' (which recommends adopting the simplest design solution for protocols and not to multiply complexity beyond what is necessary. Previous work has shown that simple protocols with severe constrains can still enable ``rich'' applications. For example, in situations where mobile users cannot access data services (e.g., due to services not being offered in that location or due to unaffordability): the users can access services through short messaging service (SMS) and voice services. In scenarios where the network is congested, users can even exploit asynchronous voice messages in contrast to live voice calls \cite{heimerl2009message}.  


\subsubsection{Sustainable Approximate Networking}

While ICT admittedly has many benefits, the unthoughtful use of technology can lead to unintended harmful side effects (e.g., when the society becomes overly reliant on technology, it becomes too reliant on it, and fails to function when technology is disrupted). As we chart out the approximate networking ecosystem, it will be a good time to base approximate networking on the strong architectural foundations of ``benign computing'' \cite{raghavan2015abstraction}, which is focused on minimizing the harmful side effects of technology. In this regard, we can focus on making our approximation networking solutions scale out, fail well, have open design at every level of its structure \cite{raghavan2015abstraction}. 

  


\section{Leveraging Limits: Doing More With Less}

\vspace{1mm}
``\textit{The impediment to action advances action. What stands in the way becomes the way.}''---Marcus Aurelius.
\vspace{1mm}

Limits in networking are usually considered as artifacts that constrain performance. But this does not necessarily be the case. We know that many expressive mediums (such as poetry/ paintings) deal with strict constraints, and the beauty and elegance of such media is in how these constraints are managed. The limits in networking can also be viewed analogously---the challenge is to develop approximations that can deal with the limits in liberating ways. 

Throwing more resources at a problem is costly and often can lead to sloppy solutions. In fact, it is well known that sometimes removing features can actually improve performance (cf. \textit{Braess' paradox} \cite{steinberg1983prevalence}). Braess' paradox can manifest itself in transportation networks when opening up new roads can counterintuitively deteriorate traffic conditions while closing down roads can sometimes improve traffic conditions. In the context of networking, it has been shown that counterintuitively the overall efficiency can improve by using a worse service \cite{mittal2013improve}. 
 
Doing more with less is not only desirable but will become imperative in the undeveloping future (as the economies of today's advanced countries become stagnate and face growing resource constraints). Fortunately, even with seriously deficient infrastructure, approximate solutions can be remarkably useful. Constrained environments can lead towards lean simple ingenuous ideas that can bypass the liability introduced by the constraint; in certain situations, by looking at the problem in the light of restricted resources, a better lean solution may be envisioned that can advance the state of the art more generally (i.e., even for situations sans the constraint). For example, the abundance of resources can mask inefficient design; while additional resources can be used to allay performance bottlenecks, under tight constraints, the possibility of improving implementation's efficiency becomes attractive since it can provide improved performance even with the bottlenecked scarce resources \cite{varghese2010network}.


The resource constraints can also unfetter a ``Jugaad''\footnote{Jugaad is a Hindi/ Urdu work used in the Indian subcontinent for a street-smart hack that can do the job. Jugaad, which can be translated as bricolage, typically involved some ingenous gaming of the system.} hacker mentality that can lead to novel technical solutions \cite{radjou2012jugaad}. To illustrate how Jugaad-based thinking can generally advance the state of the art, we highlight how IEEE 802.11 (originally a wireless local area networking standard) was discovered as a technically and economically viable long-distance communication technology by researchers who were driven by the desire to use the low-cost off-the-shelf 802.11 network interface cards (NICs) in constrained settings. 
  

Approximate networking can also lead to ``disruptive innovations'' \cite{christensen2013innovator}. According to Clayton Christensen's ``disruptive innovation'' theory \cite{christensen2013innovator}, disruptive innovations typically start as cost-efficient lower-end technologies (let's call them approximate technologies) that do not necessarily meet all the needs of their mainstream customers. For example, consider that Wikipedia initially started as an approximate Britannica Encarta, Skype as an approximate telephone service, and WhatsApp as an SMS service. The users adopt them in droves due to their cost efficiency and these systems often evolve enough eventually to displace high-end ``ideal'' reigning technologies. The recent uptake of various ``over the top'' messaging and calling services (such as WhatsApp and Skype) demonstrate the disruptive potential of these approximate networking applications.


\section{Conclusions}

\vspace{1mm}
``\textit{One cannot alter a condition with the same mindset that created it in the first place.}''---Albert Einstein
\vspace{1mm}

The deep-rooted reliance on infrastructure---which itself depends on many exogenously sourced depletable resources (such as energy and materials)---makes the modern society vulnerable to a disruptive collapse when resources become less available. To cope up with such a likely eventuality---in which the world will be burdened by fundamental limits that will globally affect developing and developed countries alike---we have proposed the idea of ``approximate networking''. Approximate networking is based on the idea that coping with such a world burdened with limits will require us to adopt context-specific tradeoffs to provide ``good enough'' service. In this paper, we have provided some basic building blocks for approximate networking solutions for networks in the LIMITS environment. Determining what these context-appropriate tradeoffs in different LIMITS settings will look like is an important open issue and needs more attention. 

\bibliographystyle{ieeetr}

\bibliography{limits}

\end{document}